\documentclass{article}
\usepackage[english]{babel}
\usepackage[utf8]{inputenc}
\usepackage[a4paper, total={6in, 9in}]{geometry}
\usepackage{graphicx}
\usepackage[skip=0.5pt]{subcaption}
\usepackage[width=.9\textwidth]{caption}
\usepackage{amsmath}
\usepackage{multirow}

\usepackage[numbers]{natbib}
\title{The temporal concentration of travel demand in an urban transport network}
\author{Carmen Cabrera-Arnau$^{1,2,*}$, Ng Liang Wei$^{2}$, Howard Wong$^{2,3}$, Chen Zhong$^{2}$}

\begin{document}

\maketitle

\begin{center}
$^1$ Department of Geography and Planning, University of Liverpool, London, UK\\
$^2$ Centre for Advanced Spatial Analysis, University College London, London, UK \\
$^3$ Transport for London, London, UK\\
$^{*}$Corresponding author: \tt{c.cabrera-arnau@liverpool.ac.uk} 
\end{center}

\begin{abstract} 
Suppose $A$ and $B$ are two stations within the mass rapid transit network of a city. Both stations see approximately the same average daily number of passengers entering and exiting their gates. However, passengers are evenly distributed at $A$, whereas activity is concentrated mainly during peak hours at $B$. Although the daily travel demand is the same for both stations, $B$ requires more resources since the number of vehicles, station dimensions and staffing level must be tailored to meet the demands of peak hours. This hypothetical scenario underscores the need to quantify the concentration of travel demand for optimising resource allocation and planning efficiency in an urban transport network. To this end, we introduce a novel metric for assessing the temporal concentration of travel demand at different locations in a generic transport network. Our approach is validated using granular data sourced from smart travel cards, encompassing 272 London Underground (LU) stations. Additionally, we present a methodological framework based on Random Forests to identify attributes of the locations of interest within the transport network that contribute to varying levels of temporal concentration of travel demand. Our case study unveils that LU stations located in areas characterised by low residential, retail, and employment density, predominantly situated in outer London, exhibit the most pronounced temporal concentration of travel demand. Conversely, within inner London, stations servicing high-density employment zones, especially around the City of London, experience a greater temporal concentration of travel demand compared to those catering to commercial and residential districts, typically situated in West London.

\end{abstract}


\section{Introduction}

\subsection{The spatiotemporal heterogeneity of travel demand}

The proportion of the world’s population living in urban areas is projected to increase from 55\% in 2018 to almost 70\% by 2050 \cite{Urbanisation18}. This growth will place pressure on urban infrastructures, particularly the transportation systems needed to support the daily movement of urban populations. Mass rapid transit (MRT) systems, also known as subway or metro networks, are frequently seen as the most efficient transportation means in large urban areas \citep{Rodrigue20} and enhancing urban mobility through the construction or expansion of such systems is therefore key for cities to manage urban growth while improving the quality of life of their people \citep{Enright20}.

One of the challenges in the planning and management of MRT systems and of other transport modes in the urban transport network is that travel demand exhibits spatial and temporal heterogeneity \citep{Rodrigue20}. The urban space can be conceptualised as a medium for different activities taking place at different locations and at different times, and the separation of activities over space and time generates the movement of people. This separation does not occur randomly; rather, it is influenced by the urban form and function \citep{Rodrigue20} and dictated by cyclical repetition of routinised processes arising from institutional or personal commitments \citep{Mulicek15}. An example is the home-to-work commute from residential areas to work locations in the morning and the reverse in the evening on working days. As a consequence, there is a spatial and temporal aspect to the heterogeneity in travel demand, where the demand experienced by stations and routes depends on the location and the time of day. 

The variability of travel demand across the urban transport network can have important operational and financial implications, and affect user experience. The locations and routes within the network that experience a high temporal concentration of travel demand are potential vulnerabilities \citep{Jiang18} since disruption at peak time can affect a large number of passengers, even when the total daily volume of passengers is relatively low. At the same time, crowdedness during peak hours creates discomfort for commuters \citep{Tirachini17}, which can increase the chances of disruption at these locations. Locations where travel demand is highly concentrated also incur higher operational costs and require more public subsidies \citep{Horcher21} since these resources need to be tailored to meet the demands of peak hours.

\subsection{Spatiotemporal patterns of urban mobility}

Empirical research into the spatiotemporal patterns of mobility in large urban agglomerations has been an active research area since the first surveys of household transportation usage were carried out in the 1960s \citep{Manley18}. With the advancement of technologies that can trace people's movements in urban areas with greater accuracy and granularity, richer travel data that allows for more insightful analysis of urban mobility patterns have emerged (for example, see \citep{Yan14, Yin17, Yang18, Ponce-Lopez21, Wang22, Rowe23}). Smart card data is a particularly useful data source. It is recorded by automatic fare collection systems on public transport networks and it can capture the location and time of entries and exits into the networks. 

In the past two decades, there have been multiple research works that use smart card data to study urban mobility patterns on MRT systems \citep{Roth11, Maeda19, Yang19, Tang20, Nilufer21, Zhang21, Cabrera-Arnau23}. Generally, the observed mobility patterns from smart travel card data display daily and weekly regularities. In addition, these patterns show notable similarities with traditional data sources indirectly related to urban activity, such as recorded crimes and road accidents \citep{PrietoCuriel23_weeklycrime, Cabrera-Arnau20}. This alignment underscores the potential for smart travel card data to offer valuable insights into urban mobility trends, providing a robust foundation for informed urban planning and policy decisions.


Beneath the temporal regularity in the level of travel demand at the aggregate level, it has also been shown that there are variations at different points of the transport network influenced by location-based characteristics \citep{Zhang14, Jun15, Tu18, He19, Du22}. It should be highlighted that all the studies cited in the previous sentence focused on the volume of travel demand and not the temporal concentration of travel demand, though some acknowledged the latter and performed separate analyses using the volume of demand at different times of the day as the dependent variable. Crucially, these studies established that the demand for travel can be influenced by access to opportunities and services at different locations, the socioeconomic factors of the area, transit-related factors, and the structure of the transport network. Whether the same factors that influence the volume of demand similarly affect the temporal concentration of demand remains an open question that can be investigated once a measure for the latter has been formulated.

\subsection{Main contribution}



The contribution of this paper is chiefly methodological. Firstly, we introduce a new metric to quantify the temporal concentration of travel demand at different locations within an urban transport network. Data sourced from smart travel cards encompassing 272 London Underground (LU) stations is used to validate our approach. For this particular case study, we define travel demand as the volume of passengers entering and (or) exiting a station at a given time of the day. We use the term throughput to refer to the sum of entries and exits at a station. We then propose a methodology to identify the most important features that lead to various levels of temporal concentration of travel demand at different locations within the transport network. Our methodology also involves a clustering process whereby locations that share similar features are grouped together, revealing some spatial patterns in the temporal concentration of travel demand. By leveraging several data sources, we find that for the LU transport network, station features with the highest impact on travel demand concentration are often related to the urban function of the area surrounding each station. 

Our findings provide insights for policy discussions and interventions. In particular, quantifying the temporal concentration may inform the daily operation of public transport. For instance, our methodology can be used by public transport operators as a tool to optimise workforce distribution through the urban RMT system. Moreover, our methodology can help understand the effects of interventions altering the urban dynamics of a specific area on the concentration of travel demand at locations within that vicinity.

\section{Data}

We should emphasise that the focus of this paper is on the proposed methodology and the analysis of data from the London Underground (LU) network is used as a means for validation. Below we describe the specific data sets for London, however, similar data sets could be obtained for other cities or other transport systems and analogous results could be generated based on the same methods that we describe below.

\subsection{Travel data}

The urban transport network chosen to validate the methodological approach proposed here is the London Underground. With a history dating back to 1863, LU is the world’s oldest underground MRT system \citep{TfLa}. Today, it operates 11 lines spanning 402km, serving 272 stations and carrying up to five million passenger journeys a day \citep{TfLb}. Data on the journeys on London’s rail network for the period from January 2020 to February 2020 and May to August 2021 is provided by Transport for London (TfL). Each record contains the following information: i) calendar date of entry, ii) time of entry at 15-min intervals (e.g., 00:00 to 00:15, 00:15 to 00:30, and so on), iii) entry and exit stations, iv) number of journeys made between entry and exit stations in each interval, v) minimum, maximum, average and standard deviation of travel time.

If there are less than five journeys between any pair of stations in any 15-min interval, this is not reported in the data. While this might result in an under-representation of activity between certain stations, this is a limitation in the accuracy of the data imposed by the data holders in order to protect the privacy of the passengers. Additionally, the following records are removed from the dataset: i) records with entry time falling outside the operating hours of 04:30 to 02:00 the next day (these account for 0.02\% of all demand), ii) records where the entry and exit stations are the same (these account for 0.002\% of all demand) and iii) records falling on bank holiday weekdays in the UK, as the travel patterns are different from a typical weekday.

\begin{figure}[ht]
    \centering
    \includegraphics[width=\textwidth]{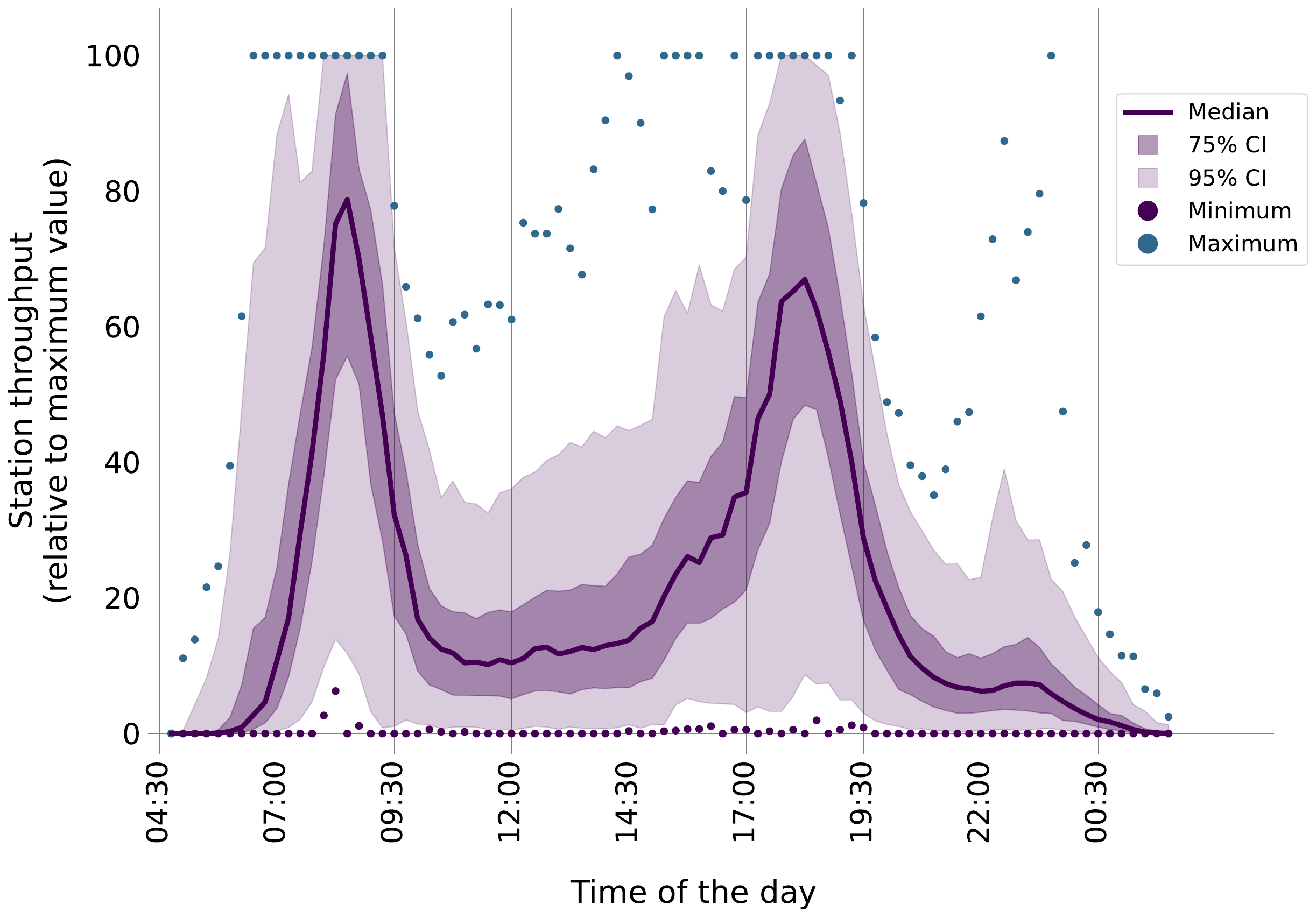}
    \caption{Evolution of the total travel demand on a typical weekday, measured as the throughput at several 
    London Underground stations.}
    \label{fig:travel_demand}
\end{figure}

As mentioned above, for this case study, we define travel demand as the volume of passengers entering and exiting a station at a given time of the day. Note that this definition might vary depending on the type of urban transport network under study. The subsequent analysis focuses on weekdays only, although a similar analysis could be applied to weekend days. Figure \ref{fig:travel_demand} shows the sum of entries and exits, i.e. the throughput, over a typical weekday. Travel demand data for a typical weekday is obtained by averaging the travel demand data for each time interval across all the days considered in the time period that the data is available. On the vertical axis, the throughput is normalised so that 100 corresponds to the maximum demand registered throughout the day at each station. The red solid line represents the average demand across stations, while the shaded region corresponds to the CI. Travel demand for an average weekday exhibits two peaks of similar height, one in the morning and one in the evening, reflecting an increase in demand most likely driven by the need to travel to or from work. 

\subsection{Other data}

We use additional sources of data for our analysis. Firstly, the geographical boundaries of Local Authority Districts (LADs) and Lower Super Output Areas (LSOAs), available for download as shapefiles from \citep{OpenGeographyPortal}, and the geographical boundaries of London’s Central Activities Zone (CAZ), also available for download as a shapefile from \citep{LondonDatastore}. We also use land use data at the LAD-level, downloaded from \citep{LandUseData} and data on population density and number of jobs at the LSOA-level, available at \citep{ONSwebsite}. Finally, we use data on National Public Transport Access Nodes (NAPTANs) locations from \citep{NaPTAN}.

\section{Measuring the temporal concentration of travel demand}

For the subsequent parts of the analysis, we introduce the following notation. Each of the $N$ locations of interest in the urban transport network is symbolised by $L^i$, with $i=1,..., N$. The travel demand at location $L^i$ is denoted by $Y^i$, which in our case study corresponds to the sum of the number of passengers entering and exiting station $L_i$. We conceptualise $Y^i$ as a time series, where we regard $Y^i$ as a collection of data points indexed in time order, so $Y^i = \{Y^i_t: t \in T\}$, where $T$ symbolises the index set. The time $t$ is parametrised so that it only takes integer values in the parameter set $T$. In our case study, we are interested in measuring the temporal concentration of travel demand on an average weekday, so $T =\{1,...,86\}$. The time index represents specific 15-minute time intervals through the operating hours of the LU network, starting at 4:30 am and ending at 2:00 am on the next calendar day. For simplicity, we refer to the operating hours period simply as ``day". For example, $t=1$ represents the first time interval of the day, i.e. from 4:30 to 4:45 am. Similarly, $t=5$ would represent the 5th time interval, i.e. from 5:30 to 5:45 am, and $t=86$ would represent the last time interval of the day, from 1:30 to 2:00 am (technically, the next calendar day). 

To measure the temporal concentration of travel demand at location $L^i$ we use the Gini index of the travel demand $Y^i$. The Gini index is a statistical index commonly used in social sciences to measure the variability in the distribution of a positive random variable \citep{Giorgi17}. It is typically used to quantify inequalities in the distribution of wealth among individuals or regions within a country. However, its definition is general enough that it can also be used in other fields of application. For example, in the area of urban transport, the Gini index has been used by H\"{o}rcher and Graham \citep{Horcher21} to measure demand imbalance on a mass rapid transit system, where they focused on the spatial aspect of demand concentration in terms of the demand distribution along the length of the metro lines. The Gini index has also been used in various disciplines to measure the temporal concentration of different phenomena. For example, it was used by Sang\"{u}esa et al. \citep{Sanguesa18} to quantify the concentration of rainfall over a year, while Vergori and Arima \citep{Vergori22} used it to compare the seasonality of international tourist arrivals via different transport modes. Here we use it to measure the temporal concentration of travel demand at different locations of an urban transport network, following an approach similar to that proposed by Prieto Curiel et al. in \citep{PrietoCuriel23_weeklycrime}, where the authors used the Gini index to measure the temporal concentration of crime.

The Gini index at location $L^i$ is denoted by $G^i$ and is defined as:
\begin{equation}\label{Gini}
    G^i = \frac{\sum_{t_1=1}^{|T|}\sum_{t_2=1}^{|T|}|Y^i_{t_1}-Y^i_{t_2}|}{2\sum_{t_1=1}^{|T|}\sum_{t_2=1}^{|T|} Y^i_{t_2}}
\end{equation}
where $|T|$ symbolises the number of time intervals through the time period of interest (i.e. a working day, a week, etc.) and $|Y^i_{t_1}-Y^i_{t_2}|$ symbolises the absolute value of the difference between the value of $Y^i$ at two time intervals. The Gini index can therefore take values between 0 and 1. A value of $G^i=0$ would correspond to a situation where the travel demand at location $L^i$ is perfectly evenly distributed across the time intervals under consideration. By contrast, a value of $G^i$ close to $1$ would correspond to a situation where all the travel demand at location $L^i$ is concentrated in one or just a few of the time intervals.

\subsection{Testing the statistical significance of the temporal concentration of travel demand}

To test the statistical significance of the temporal concentration of travel demand, we apply the following steps to each station:

\begin{itemize}
    \item Firstly, we obtain the 95\% confidence interval for the Gini index that would arise if travel demand was uniformly distributed across all the time intervals in the period of interest, which we denote by $G^{i(U)}$ to emphasise that it arises from a uniform distribution. To do this:
    \begin{itemize}
        \item We generate uniformly distributed random samples of $Y^{i(j)}$ where the $j$ in brackets refers to the $j$th sample and $j=1,..., 1000$. Each random sample $Y^{i(j)}$ is generated so that the total travel demand is equal to the observed total daily travel demand, i.e. $\sum_{t=1}^{|T|}Y^{i(j)}_t = \sum_{t=1}^{|T|}Y^i_t$, with $j=1,...,1000$.
        \item For each sample, we compute the Gini index according to equation \eqref{Gini} and denote it by $G^{i(j)}$.
        \item The 95\% confidence interval for $G^{i(U)}$ is defined by a lower and an upper bound, which can be obtained as the 2.5 and 97.5 percentiles of the range of values of $G^{i(j)}$ obtained from all the samples.
    \end{itemize}
    \item Then, we obtain a $p$-value for the Gini index by iterating the following procedure $1000$ times.
    \begin{itemize}
        \item We generate a bootstrap sample of size $|T|$ from the observed $Y^i$ and call this sample $Y^{i(k)}$, with $k=1,...,1000$. 
        \item We then compute the Gini index of $Y^{i(k)}$ and denote it by $G^{i(k)}$.
        \item We check whether $G^{i(k)}$ falls within the 95\% confidence interval for $G^{i(U)}$. 
    \end{itemize}
    To obtain the $p$-value, we count the proportion of times that $G^{i(k)}$ falls within the 95\% confidence interval for $G^{i(U)}$. The $p$-value gives the probability that the Gini index would take the observed value if the data was randomly distributed. 
\end{itemize}

\subsection{The spatial variability of the temporal concentration of travel demand}

\begin{figure}
    \centering
    \includegraphics[width=\textwidth]{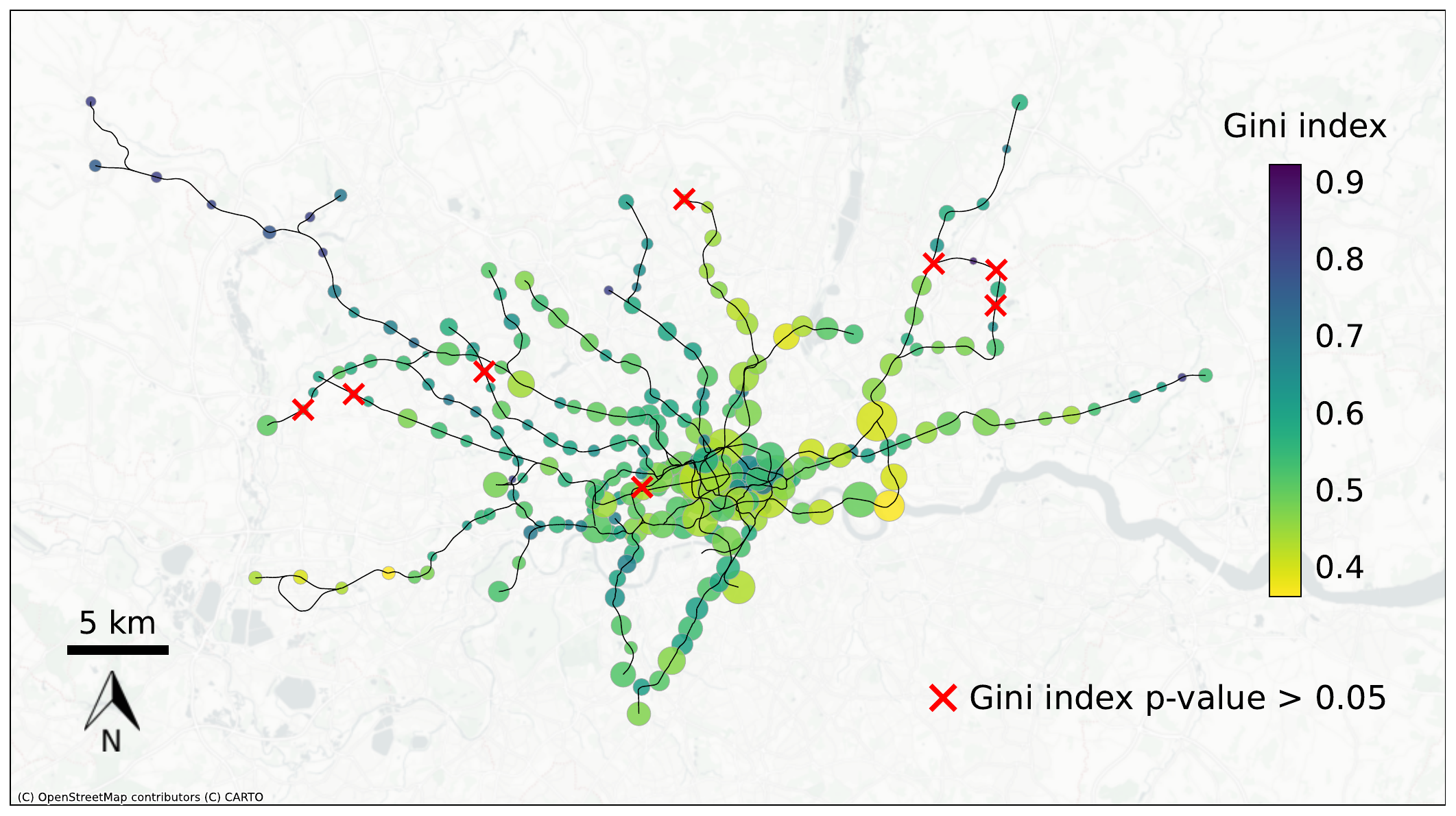}
    \caption{Concentration of travel demand across the London Underground network, as measured by the Gini index of the throughput at each station on a typical weekday. The size of the circles for each station represents the volume of the throughput.}
    \label{fig:map_gini}
\end{figure}

Figure \ref{fig:map_gini} shows the observed values of the temporal concentration of travel demand at the stations in the LU network. For the map, the travel demand is computed as the throughput at each station. There are spatial patterns in the temporal concentration of travel demand, whereby some stations in peripheral areas tend to display higher values than their inner-city counterparts. Stations where the Gini index had an associated $p$-value of more than 0.05 are marked with a cross and discarded from the analysis. We explore the drivers of travel demand in the remainder of the paper.

\section{Factors influencing the temporal concentration of travel demand}

The rest of the analysis focuses on identifying attributes of the locations of interest within the urban transport network that contribute to varying levels of temporal concentration of travel demand. We illustrate the methodology for the case study of the LU network, although this methodological framework is easily generalisable to other urban networks as long as enough data is available. 

\subsection{Feature selection}

Random Forest (RF) models are used to identify the LU station features with the strongest influence on the values of $G^i$. Table \ref{tab:features} summarises the features that are considered in this paper. These features are chosen based on prior related literature \citep{Zhang14, Jun15, Tu18, He19, Du22} as well as the availability of data. Features are derived using a service area of a 1 km radius around each station. For features where the original data is already spatially aggregated, a weighted average is computed using the area of intersection between the service area and the original spatial unit of aggregation. Two limitations are highlighted here due to the lack of availability of better data. Firstly, land use data is aggregated at the Local Authority District (LAD) level. LADs are a level of subnational division of England used for the purposes of local government. However, LADs tend to be larger than the stations’ service areas considered in this work. Secondly, the data for the station features corresponds to different years, as shown in Table \ref{tab:features}.

\begin{table} [ht]
    \caption{Station features considered in the analysis. CAZ stands for Central Activities Zone.}
    \vspace{-.2cm}
    \label{tab:features}
    \hspace{-.2cm}\includegraphics[width=1.05\textwidth]{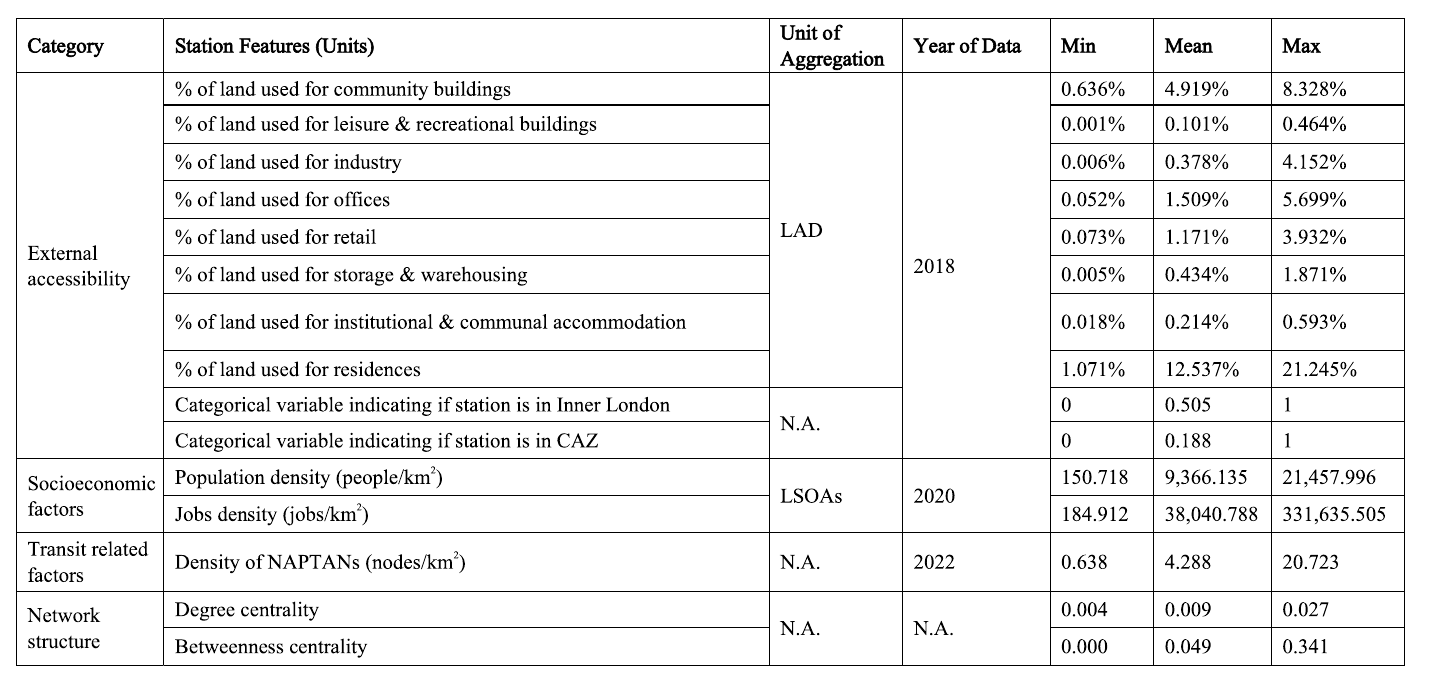}
\end{table}


It can be shown that some of the features exhibit high correlation 
and it is possible to reduce the dimensions using, for example, Principal Component Analysis, but a decision not to do so is made for two reasons. Firstly, while some of the features are correlated, they may capture different drivers for the temporal concentration of demand. For example, retail outlets may generate employment opportunities typically associated with the concentration of travel demand during peak hours, but also serve as points of interest that generate travel demand during off-peak hours, especially in tourist areas. Secondly, the station features considered here are of a different nature, which may pose a challenge in the interpretation of principal components. Hence, as the performance of RF and K-Means clustering (used in the next step) is not known to be adversely affected by collinearity in input features, all the station features are used for subsequent analyses directly.

Another issue related to the correlation of station features is the approach for determining feature importance. A common method is to use the mean decrease in impurity to measure how effective a feature is at reducing the variance when creating decision trees in RF, but this approach inflates the importance of continuous variables and high-cardinality variables \cite{Strobl2007-yr}. Another method proposed by \cite{Breiman2001} is based on permutation importance, where the feature importance is measured by the increase in a model’s prediction error after permuting a feature – the higher the increase, the more important the feature. However, \cite{Strobl2008-io} showed that this method over-estimates the importance of correlated features. Building upon the same idea, \cite{Parr2018} suggested basing the importance of a feature on its impact on the model accuracy after dropping it from the model totally, instead of permuting it. Even though this method incurs higher computational cost as the model needs to be re-trained every time a feature is dropped (ibid.), this method is more appropriate in this paper, given the underlying data structure

Table \ref{tab:hyperparameters} shows the search space for the hyperparameter tuning of the RF models to optimise their performance in this dissertation, based on the recommendations of Probst, Wright and Boulesteix in \cite{Probst19}. The authors of this article also argued that the number of trees cannot be seen as a tuning hyperparameter since there is no trade-off in the accuracy of the model when more trees are used, other than the increased computational requirements. Nonetheless, \cite{Lunetta2004} proposed using thousands of trees to obtain stable estimates of feature importance. Hence, the RF models are trained using 1000 trees. In addition, the hyperparameters are tuned using the Mean Squared Error (MSE) as the scoring criterion.

\begin{table} [ht]
    \caption{Hyperparameter search}
    \label{tab:hyperparameters}
    \vspace{-.5cm}\centering
    \includegraphics[width=0.9\textwidth]{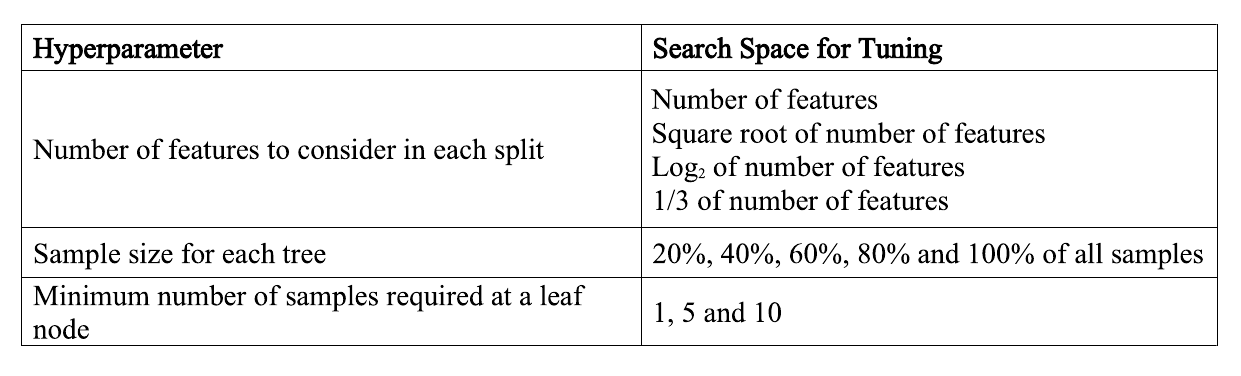}
\end{table}
	
After finding the optimal hyperparameters, feature selection is performed as follows:
\begin{enumerate}
    \item Starting with all features, train the RF model with 75\% of the stations, test the accuracy with the remaining stations using the MSE, and rank the features by their importance. Remove the least important feature.
    \item Repeat Step (1) until there is only one feature left.
    \item Select the set of features with the lowest root of MSE.
\end{enumerate}

The selected features are population density, job density, the proportion of land use dedicated to retail, the proportion of land use dedicated to institutional and communal accommodation, and the betweenness centrality of the stations in the transport network.

\subsection{Station classification into clusters based on the selected features}

Stations are classified based on the similarity of the selected features using the K-Means clustering algorithm with 500 iterations. Features are first standardised. To determine the optimal number of clusters, the silhouette score (SS) is initially computed for 2 to 10 clusters. Based on the SS method, the optimal number of clusters is 2. However, our comprehensive understanding of the intricacies of London's urban dynamics, coupled with visual analysis of the clusters on a map, led us to set the number of clusters to 4. This decision stems from our domain expertise, aiming to capture the diverse urban patterns effectively and to improve the interpretability of our results. 

\subsection{Interpretation of clusters}

\begin{figure}
    \centering
    \fbox{\includegraphics[width=\textwidth]{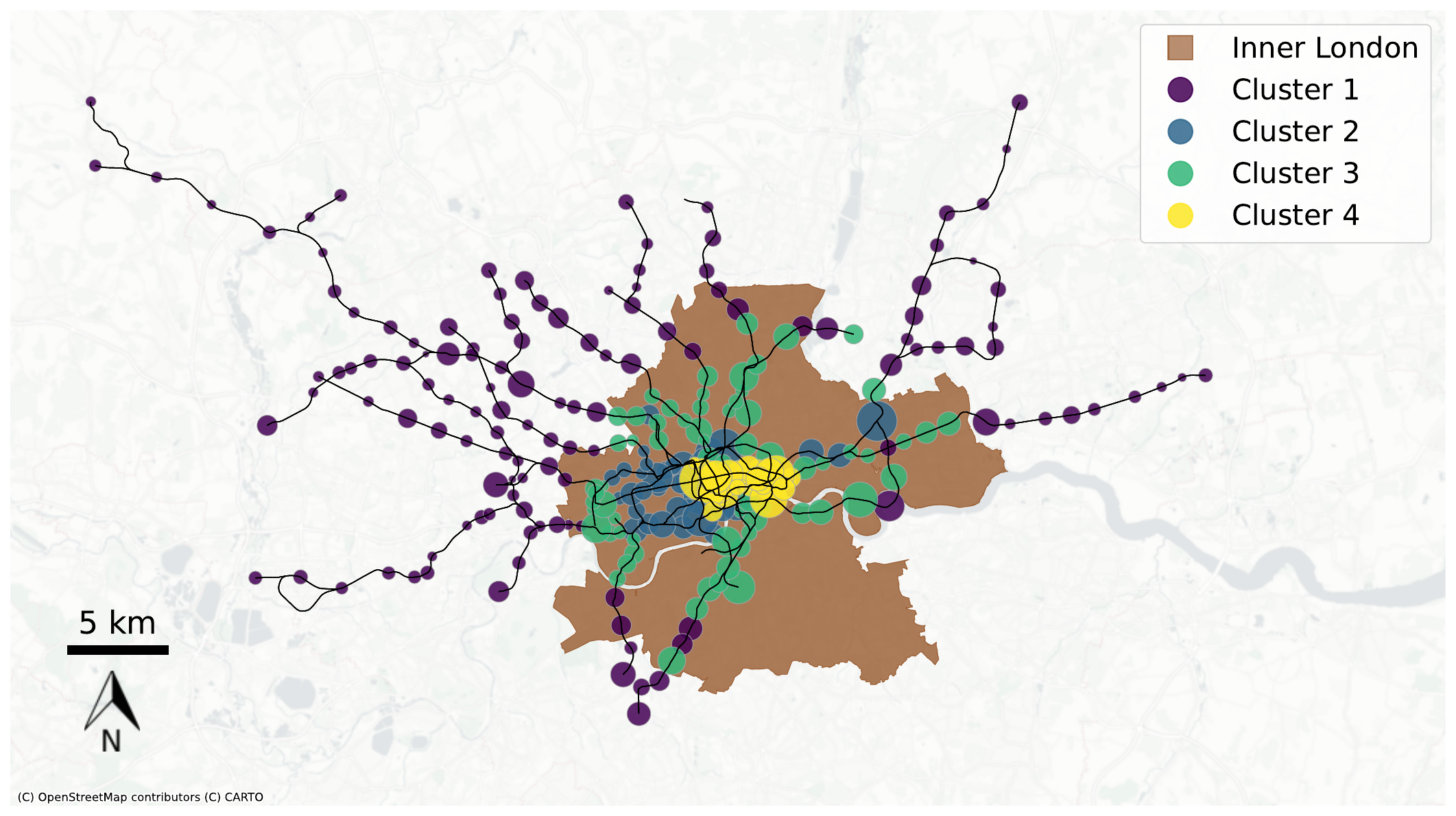}}
    \caption{Geographical location of each cluster. The shaded area in the map represents inner London.}
    \label{fig:cluster_location}
\end{figure}

\begin{figure}
    \centering
    \fbox{\includegraphics[width=\textwidth]{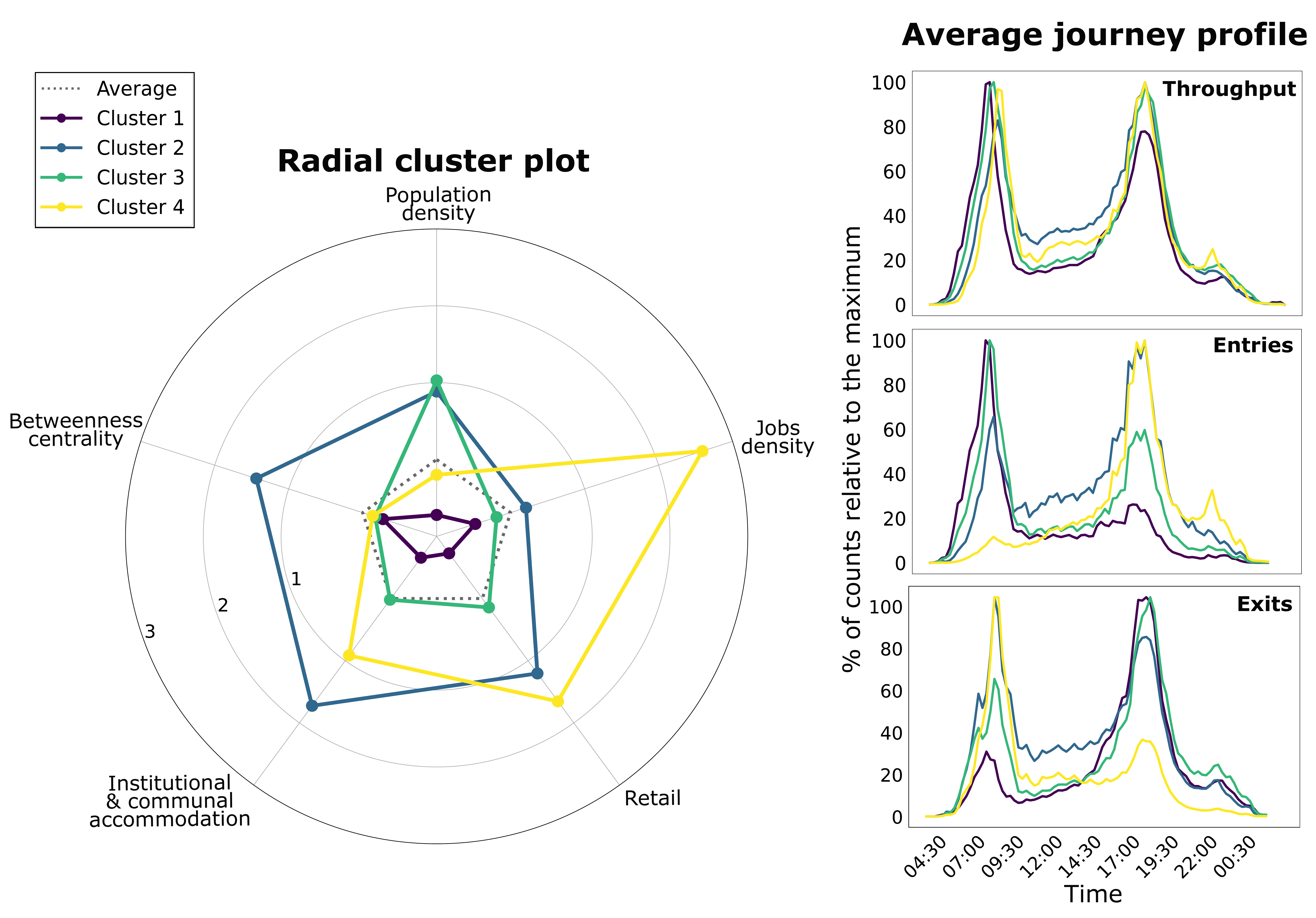}}
    \caption{Radial cluster plot and average journey profile of throughput, entries and exits for each cluster.}
    \label{fig:cluster_interpret}
\end{figure}

Figure \ref{fig:cluster_location} shows where the clusters of stations are geographically located. Clusters 2, 3 and 4 are within inner London, and Cluster 1 corresponds to outer London. The radial cluster plot in Figure \ref{fig:cluster_interpret} shows the profile of the resulting groups of stations. On the right-hand side of the same Figure, we include the average journey profile for the throughput, the entries and the exits at each station. Based on the information displayed in Figures \ref{fig:cluster_location} and \ref{fig:cluster_interpret}, we give a description of each cluster below.

\textbf{Cluster 1} corresponds to the stations in outer London. According to the radial cluster plot from Figure \ref{fig:cluster_interpret}, this cluster is characterised by stations in areas with low population density, low job density and low proportion of land use dedicated to retail or to institutional and communal accommodation. The average journey profile for the stations in this cluster reveals that most entries are concentrated during the morning rush hour, while most exits take place during the evening rush hour. Hence, Cluster 1 represents mostly residential suburbs outside the inner core of the city, with lower than average population density.

\textbf{Cluster 2} corresponds to stations in central areas of inner London, characterised by relatively high population density, betweenness centrality and proportion of land use dedicated to institutional and communal accommodation and to retail. The job density is slightly above average. Therefore, Cluster 2 represents the residential areas in central London. In contrast with Cluster 1, Cluster 2 sees most entries in the evening peak hour. This reflects the fact that the passengers might come to these areas during the day for work or leisure, but the majority of those who do not live in Cluster 2 get back to their place of residence in the evening. The average profile of exits follows a complementary pattern, i.e. there are more exits in the morning than in the evening rush hour, which shows that public transport users are attracted to spend the day in the areas surrounding the stations in Cluster 2.

All the features are around the average level in \textbf{Cluster 3}, except for population density, which is higher than average. Based on its location, Cluster 3 represents the peripheral stations in inner London, where housing is in high demand among city dwellers given that its cost is typically lower than in the very central areas while retaining the proximity to many of the amenities present in central areas. Similar to Cluster 1, most entries are concentrated during the morning rush hour, although there are many more entries in the evening rush hour than in the case of Cluster 1. This reflects the higher job density and the higher number of retail outlets that are available in Cluster 3 but not in Cluster 1. Analogously, most exits are concentrated during the evening rush hour like in Cluster 1, but there are more exits in the morning rush hour than in the case of Cluster 1, reflecting the in-flow of commuters from other parts of the city that go to work in the proximity of the stations situated in Cluster 2.

High job density and a high proportion of land use dedicated to retail are the most remarkable features in \textbf{Cluster 4}, which corresponds to the CBD of London. In this part of the city, there is also a higher than average proportion of land use for institutional and communal accommodation, and slightly lower than average population density. The average journey profile for the stations in this cluster is similar to those in Cluster 2, except the patterns are more exaggerated due to the fact that the CBD is very much a job-centred area. Most entries are concentrated during the evening rush hour, while most exits take place during the morning rush hour, reflecting the fact that the majority of people travel to Cluster 4 for work and once the work day is over, they leave the area to go back to their place of residence.

\begin{figure}
    \centering
    \includegraphics[width=1\textwidth]{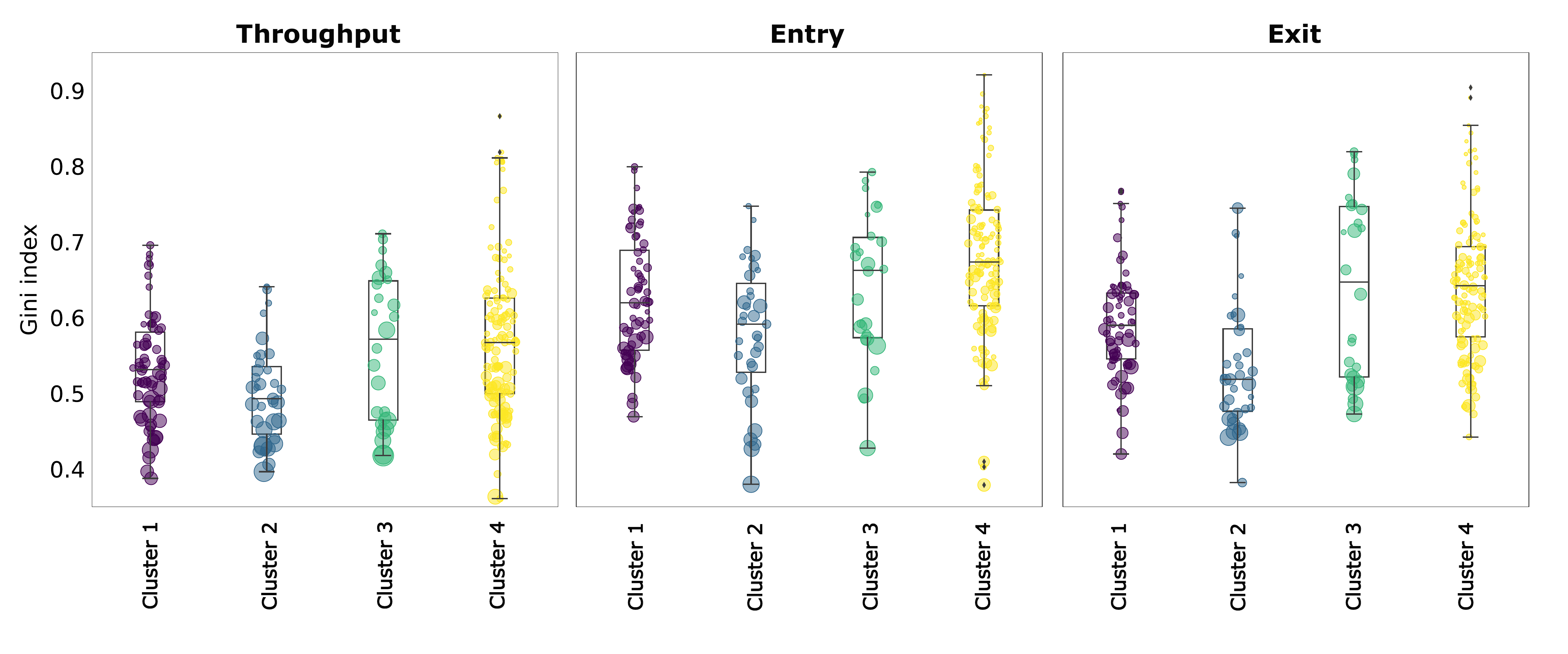}
    \caption{Boxplots representing the values of the Gini index for stations grouped by cluster, for throughput, entries and exits.}
    \label{fig:boxplots}
\end{figure}

In Figure \ref{fig:boxplots}, we display boxplots showing the range of values of the Gini index for stations grouped by cluster. We include values of the Gini index at each station computed for the travel demand according to the throughput, the number of entries and the number of exits. We observe that Cluster 2 is consistently the one with the lowest values of the Gini index. This reflects the fact that the stations belonging to Cluster 2 are surrounded by areas of mixed purpose, characterised by the presence of places of residence, jobs and retail above average levels, as shown in the radial plot from \ref{fig:cluster_interpret}. 

Cluster 1 has generally higher values of the Gini index than Cluster 2. This is a testament to the fact that the stations in this cluster are located mostly in residential suburbs. Therefore, passengers enter their local station early in the day at the start of their morning commute and exit late at the end of their evening commute. Cluster 1 is also characterised by a relatively low variance, reflecting the fact that the travel behaviour in these stations is comparatively homogeneous.

The level of temporal concentration in stations from Cluster 3 is generally similar to the level observed for stations in Cluster 1. This suggests a similar travel behaviour from passengers using stations in Clusters 1 and 3, which are both characterised by a higher than average population density and their location is not as central as Clusters 2 and 4.

Cluster 4 registers high values of the Gini index, especially for the entries. This can be attributed to the fact that Cluster 4 corresponds to stations in the CBD, where people travel to early in the day for work and they leave in a more staggered fashion as they finish their working day. The values of the Gini index for the stations in Cluster 4 are characterised by their high variability, suggesting a heterogeneous travel behaviour across the stations belonging to this cluster.

\section{Discussion}

This paper set out to examine the spatiotemporal variability of travel, using the LU network as a case study and with a particular focus on the concentration of travel demand. Firstly, the Gini index is used to measure the temporal concentration. While the demand profile can provide more complete information about the distribution of travel demand, we argue that the Gini index serves as a simple tool to initially identify or diagnose stations displaying unusual levels of temporal concentration of demand. The spatial patterns displayed by the temporal concentration of travel demand are also visually examined. A tendency is observed for peripheral stations to register higher values of the Gini index, owing to the fact that stations in central areas of the city have a more consistent and less concentrated flow of passengers entering and exiting their gates throughout the day.

Secondly, Random Forest models are used to determine the features of the area surrounding the stations that drive the emergence of different journey profiles at the daily level. We find that the concentration of travel demand is influenced mostly by the population density, density of jobs, the proportion of land dedicated to retail, the proportion of land dedicated to institutional and communal accommodation, and the betweenness centrality of the stations in the transport network. However, there might be other variables not considered here that might also play a role in determining the temporal concentration of travel demand.

Using the K-means clustering algorithm, stations are then grouped into four clusters in terms of the similarity in the selected features. Different average journey profiles can be observed in each cluster, with their associated values of the Gini index. The clusters also display clear spatial patterns that match with regions typically known for fulfilling different urban functions in the specific case study of London.

Our work raises the practical question of how can our findings help transport planners ensure the efficiency of urban transport. If the temporal concentration of travel demand in a given location is high to the point of being problematic, is there anything within a transport planner's sphere of influence to solve it? Of all the candidate features considered in this paper, the one that transport planners arguably have more control over in the short- to mid-term is the NAPTAN density, which was included to capture the effect of intermodal connectivity. However, this feature turned out to be unimportant in predicting the temporal concentration of travel demand. In retrospect, the answer to the question lies in an integrated, strategic master planning of the city so that travel demand can be spread more evenly over space and time. For London, traditionally a monocentric city \citep{Cabrera-Arnau23}, the implementation of opportunity areas to de-centralise living, working and recreational opportunities away from Central London represents a positive step in this direction \citep{GLA21}.

\section*{Acknowledgements}

This project has received funding from the European Research Council (ERC) under the European Union’s Horizon 2020 research and innovation programme (Grant Agreement No. 949670), and from ESRC under JPI Urban Europe/NSFC (Grant No. ES/T000287/1). The data analysis was done within UCL.

\bibliography{Bibliography.bib} 

\end{document}